\documentstyle[preprint,prc,aps]{revtex}
\begin{document}
\preprint{DOE/ER/40561-112/INT93-00-37}
\title{Hanbury-Brown--Twiss Analysis in a Solvable Model}
\author{
G.F. Bertsch$^{(1,2)}$\thanks{Internet
 address: bertsch@phys.washington.edu},
P. Danielewicz$^{(3)}$\thanks{Internet
address: danielewicz@nscl.nscl.msu.edu},
M. Herrmann$^{(1)}$\thanks{Internet
 address: herrmann@ben.npl.washington.edu}}
\address{
$^{(1)}$ Institute for Nuclear Theory, HN-12, University of
Washington, Seattle, WA  98195\\
$^{(2)}$ Physics Department, FM-15, University of Washington,
Seattle, WA 98195\\
$^{(3)}$ Dept. of Physics and Astronomy and National
Superconducting Cyclotron Laboratory, Michigan
State University, East Lansing, MI 48824 }
\date{September 7, 1993}
\maketitle
\begin{abstract}

The analysis of meson correlations by Hanbury-Brown--Twiss
interferometry is tested with a simple model of meson production
by resonance decay.
We derive conditions which should be satisfied in order to relate
the measured momentum correlation to the classical source size.
The Bose correlation effects are apparent in both the ratio of
meson pairs to singles and in the ratio of like to unlike pairs.
With our parameter values, we find that the single particle distribution
is too distorted by the correlation to allow a straightforward analysis
using pair correlation normalized by the singles rates. An analysis
comparing symmetrized to unsymmetrized pairs is more robust, but
nonclassical off-shell effects are important at realistic temperatures.
\end{abstract}

\narrowtext
\section{Introduction}
\label{sec:intro}

    An important observable for studying high energy collisions
is the momentum correlation between emitted mesons.
  In the
statistical limit the correlation function depends on the Bose
symmetry of the particles as in the Hanbury-Brown--Twiss (HBT) analysis
of photon correlations from stars
(for a review see \cite{Bo90}).
  Under the statistical
assumptions, the meson correlation directly measures the
size of the meson interaction zone, and thus indirectly
reflects the hadronic dynamics prior to the meson freeze-out.
In particular, the correlation between pions and between kaons
has been analyzed to give quantitative data on the mean square
source dimensions \cite{Ak87,Se92,Su93,Hu93,Ak93}.

     However, the analysis is dependent on the statistical
assumptions, raising the question of their validity under
realistic conditions of meson production.  In this work we
develop a simple model of meson production to test the HBT
analysis. Within this model we can also
 examine the concept of a classical
source.
 Before defining the model, we shall briefly review the
 HBT analysis to establish the notation.  The basic observables are
the one- and two-particle differential cross sections,
$ d \sigma^{(1)} / d^3 q $ and $ d \sigma^{(2)} / d^3
q_1 d^3 q_2 $,
within some
class of events.  Writing the cross section for these events as
$ \sigma_0 $, we define the doubles-to-singles correlation
function as \cite{Gy79}

\begin{equation}\label{eq:i:1}
C^{\text{d/s}}\left(\mbox{\boldmath{$q$}}_{\text{av}},
\mbox{\boldmath{$q$}}_{\text{rel}}\right)
=
\frac{
\sigma_0 \,\frac{\displaystyle d \sigma^{(2)}}{\displaystyle
  d^3 q_1 d^3 q_2} }{
\frac{ \displaystyle d \sigma^{(1)}}{
\displaystyle d^3 q_1}\frac{\displaystyle d \sigma^{(1)}}{
 \displaystyle d^3 q_2 }}.
\end{equation}
The three-vectors $\mbox{\boldmath{$q$}}_1 $
 and $\mbox{\boldmath{$q$}}_2 $ denote the
three-momenta of the two mesons. To facilitate later on the interpretation
of the correlation function, we chose the arguments to be the average
and relative momentum,
$\mbox{\boldmath{$q$}}_{\text{av}}\equiv
 (\mbox{\boldmath{$q$}}_1 + \mbox{\boldmath{$q$}}_2)/2$
 and $\mbox{\boldmath{$q$}}_{\text{rel}}
\equiv \mbox{\boldmath{$q$}}_1 - \mbox{\boldmath{$q$}}_2$, respectively.

In the theoretical model, it will be more convenient to express
the correlation function using
decay rates rather than cross sections.  The rates are
related to cross sections by a flux factor.
In terms of
the total decay rate $W$ and the differential decay rates
$W^{\left(2\right)}(\mbox{\boldmath{$q$}}_1,
\mbox{\boldmath{$q$}}_2) = d W / d^3 q_1 d^3
q_2$
and $ W^{\left(1\right)}(\mbox{\boldmath{$q$}}) = d W /d^3 q $,
we evaluate the
doubles-to-singles correlation function as
\begin{equation}\label{eq:d/s}
C^{\text{d/s}}\left(
\mbox{\boldmath{$q$}}_{\text{av}},
\mbox{\boldmath{$q$}}_{\text{rel}}\right)
 =
\frac{W  \,W^{\left(2\right)}\left(
\mbox{\boldmath{$q$}}_1,
\mbox{\boldmath{$q$}}_2\right)}{
W^{\left(1\right)}\left(
\mbox{\boldmath{$q$}}_1\right)
W^{\left(1\right)}\left(
\mbox{\boldmath{$q$}}_2\right)} .
\end{equation}

To allow an HBT analysis of $C^{\text{d/s}}$, we have to make statistical
assumptions
(explained in the next section)
that reduce the information contained in the two-particle quantities
 to that contained in the density matrix for a single particle source.
The latter is conveniently expressed
in the Wigner
representation as the (one-particle)
phase space distribution $g^{(1)} (t,\mbox{\boldmath{$x$}},
\omega,\mbox{\boldmath{$q$}})$,
 where $\omega$ is the energy of the meson.  The connection
to the correlation function is given by Pratt's
formula \cite{Pr84},
which we will derive in the next section.  The formula reads
\widetext
\begin{equation}
\label{eq:Pratt}
C^{\text{Pratt}}(
\mbox{\boldmath{$q$}}_{\text{av}},
\mbox{\boldmath{$q$}}_{\text{rel}})
= 1 +
{\int d^4 x_1 d^4 x_2\, g^{(1)}(x_1,q_{\text{av}})g^{(1)}(x_2,q_{\text{av}})
 \cos q_{\text{rel}}
(x_1-x_2) \over \int d^4 x_1 g^{(1)}(x_1,\widetilde{q}_1)
 \int d^4 x_2 g^{(1)} (x_2,
\widetilde{q}_2) } .
\end{equation}
\narrowtext
We use a notation here with four-vectors
$x = (t,\mbox{\boldmath{$x$}}) $
and $ q = (\omega,\mbox{\boldmath{$q$}}) $.
To separate on-shell from off-shell
four-momenta we use the notation $\widetilde{q}
\equiv (\omega(\mbox{\boldmath{$q$}}),\mbox{\boldmath{$q$}})$
in the on-shell case, where
 $\omega (\mbox{\boldmath{$q$}}) \equiv
 \sqrt{\mbox{\boldmath{$q$}}^2+m_{\pi}^2}$
is the energy of a meson with momentum
 $\mbox{\boldmath{$q$}}$ and mass $m_{\pi}$.
 We have also defined the relative and
the average four-momenta
 of the two mesons, $q_{\text{av}} \equiv (\widetilde{q}_1+\widetilde{q}_2)/2$
and $ q_{\text{rel}} \equiv \widetilde{q}_1-\widetilde{q}_2$.

     To compare with theoretical expectations, it is common to
model the evolution of the interacting hadronic system using
classical transport equations \cite{Pr86,Be88,Gy89}.  With
these simulations one
can predict a classical source distribution function,
$g^{(1) \text{cl}}(t,\mbox{\boldmath{$x$}},
\sqrt{m^2+\mbox{\boldmath{$q$}}^2},\mbox{\boldmath{$q$}})$.  Note that the
energy is not an independent variable in the classical source.
Thus in principle there is not enough information to use $g^{(1) \text{cl}}$
in Eq.\ (\ref{eq:i:1}),
 since the required $q_{\text{av}}$ in that equation need not
have the energy on mass shell.  However, in favorable cases the
extrapolation off mass shell will not cause difficulties.

     We shall also find it useful to analyze the correlation data
by dividing the two-particle distribution function by the
unsymmetrized two-particle distribution. This is defined
\begin{equation}\label{eq:l/u}
C^{\text{l/u}}(\mbox{\boldmath{$q$}}_{\text{av}},
\mbox{\boldmath{$q$}}_{\text{rel}}) =
{W^{(2)}(\mbox{\boldmath{$q$}}_1,
\mbox{\boldmath{$q$}}_2) \over
W^{(2)}_u(\mbox{\boldmath{$q$}}_1,
\mbox{\boldmath{$q$}}_2) },
\end{equation}
where $W^{(2)}_u$ is the two-particle decay rate in the absence
of Bose correlations.  These  correlations are not present for pairs
of unlike mesons, so one could imagine measuring $C^{\text{l/u}}$ by
e.g.
comparing the correlation between pairs of $\pi^+$ mesons
to the correlation between $\pi^+$ and $\pi^-$.  We therefore
shall call this the like-to-unlike correlation.  However,
although the Bose interference is  absent from the unlike
decay distribution, the interaction between unlike
mesons induces stronger correlations than for charged
mesons of like sign.  Another approach to measure
$W^{(2)}_u$ is to try to extract it from the correlated data by
expressing the rate as a sum of two terms, one the
uncorrelated product of two single-particle decays, and
the other the correlation.  This has been done in some
experimental data analyses \cite{Za84,Hu93}.  We will see
in this work that
it is easier to satisfy the statistical assumptions for the
like-to-unlike ratio than for the ratio of doubles to
singles.

\section{Derivation of the Pratt formula}
\label{sec:pratt}
  The basic formula for the correlation function in terms of the
Wigner distribution of the source, Eq.\ (\ref{eq:Pratt}), was derived by Pratt
assuming an ensemble of classical currents.  We will only use
quantum mechanical amplitudes, as for example calculated from
Feynman diagrams, as it is useful to see what
assumptions are necessary to derive the formula from such
quantum mechanical expressions.

Our starting point is the  symmetrized amplitude
\begin{equation}\label{eq:1}
T_\alpha^{S}\left(q_1,q_2\right)
 =  T^{\left(2\right)}_\alpha\left(q_1,
q_2\right)
 + T^{\left(2\right)}_\alpha\left(q_2, q_1\right)
\end{equation}
that describes the production of two identical mesons with
momentum $q_1$ and $q_2$ in some
reaction.
Thus the events that contribute to the cross section $\sigma_0$ are
those that produce just two mesons.
The label $\alpha$ represents all final state variables that are
not observed, such as the number of particles and their momenta.
We separate these variables into three sets $\alpha \equiv \left(
\alpha_1 | \alpha_2 | \alpha_3 \right) $ as follows. In the unsymmetrized
amplitude $T^{\left(2\right)}_{\left(\alpha_1|\alpha_2|\alpha_3\right)}
\left(q_1, q_2\right)$
the label $\alpha_1$ denotes the variables that directly influence
the production of the meson with momentum $q_1$, the label $\alpha_2$
the ones that influence the meson with momentum $q_2$, whereas $\alpha_3$
denotes all other variables in the final state.

Up to a possible overall sign, the unsymmetrized amplitude
satisfies
\begin{equation}\label{eq:2}
T^{\left(2\right)}_{\left(
\alpha_1|\alpha_2|\alpha_3\right)}\left(q_1, q_2\right)
=
T^{\left(2\right)}_{
\left(\alpha_2|\alpha_1|\alpha_3\right)}\left(q_2, q_1\right) .
\end{equation}
The differential decay rate of the
initial state into two mesons is given by
\widetext
\begin{equation}\label{eq:3}
W_P^{\left(2\right)}\left(
\mbox{\boldmath{$q$}}_1,\mbox{\boldmath{$q$}}_2\right)
 \equiv \frac{1}{2} \sum_{\alpha} \left| T_{\alpha}^{ S}
\left(\widetilde{q}_1, \widetilde{q}_2\right)
\right|^2
(2\pi)^4\delta^4\left(P_\alpha+\widetilde{q}_1+\widetilde{q}_2-P\right).
\end{equation}
\narrowtext
Here $P$ denotes the energy-momentum vector of the
initial state. $P_\alpha$ is the energy-momentum carried away
by all unobserved degrees of freedom. The sum over $\alpha$
represents summation over discrete degrees of freedom and
integration over three-momenta, $\int d^3 p/(2\pi)^3$.
The single-particle decay rate $W^{(1)}$ and $W$ are
given by integrals of $W^{(2)}$ over one and over both momenta,
respectively.

To obtain expressions for the correlation functions
given  in Eq.\ (\ref{eq:d/s}) and Eq.\ (\ref{eq:l/u})
we first define the two-particle Wigner function
\begin{eqnarray} \label{eq:6}
&& g_P^{\left(2\right)}\left(x_1,q_1;x_2,q_2\right)
 = \frac{1}{W_{P u}}
\sum_{\alpha} \int {d^4p_1 d^4 p_2 \over (2 \pi)^8 }
\text{e}^{-i \left(p_1 x_1 + p_2 x_2\right)}
\nonumber\\
&&\times T^{\left(2\right)}_{\alpha}\left(q_1+p_1/2,
q_2+p_2/2\right)
 T^{\left(2\right) * }_{\alpha}\left(q_1-p_1/2,
q_2-p_2/2\right) \nonumber\\
&& \times (2\pi)^4  \delta^4\left(P_{\alpha} + q_1 + q_2 -P\right).
\end{eqnarray}
The notation $W_{P u}$ for the normalization  will be explained below.
The  inverse Wigner transform reads
\widetext
\begin{eqnarray} \label{eq:7}
& & \frac{1}{W_{P u}} \sum_{\alpha}
T^{\left(2\right)}_{\alpha}\left(q_1, q_2\right)
T^{\left(2\right) * }_{\alpha}\left(q_3, q_4\right)
(2\pi)^4\delta^4\left(P_{\alpha} +\sum_{i=1}^{4} q_i -P\right)
\nonumber \\
& = &  \int d^4 x_1 d^4 x_2
\text{e}^{i\left[\left(q_1-q_3\right) x_1 + \left(q_2-q_4\right) x_2\right]}
g^{\left(2\right)}_P\left(x_1,\frac{q_1+q_3}{2};x_2, \frac{q_2+q_4}{2}\right).
\end{eqnarray}
\narrowtext
Inserting (\ref{eq:7}) into (\ref{eq:3}) yields
\widetext
\begin{equation}\label{eq:8}
\frac{W^{\left(2\right)}_P \left(
\mbox{\boldmath{$q$}}_1,
 \mbox{\boldmath{$q$}}_2\right)}{W_{P u}} =
\int d^4 x_1 d^4 x_2 \nonumber
 \left[
g_P^{\left(2\right)}\left(x_1,\widetilde{q}_1;x_2, \widetilde{q}_2\right)
+
g_P^{\left(2\right)}\left(x_1,q_{\text{av}};x_2, q_{\text{av}}
\right)
\cos\left(q_{\text{rel}}\left(x_1-x_2\right)\right)
\right].
\end{equation}
\narrowtext
We can now express the like-to-unlike correlation function
 $C^{\text{l/u}}$, Eq.\ (\ref{eq:l/u}), in terms
 of two-particle Wigner functions if we
identify the unlike particle production rate with the
unsymmetrized two-particle decay rate $W^{(2)}_{P u}$
\begin{eqnarray}
 W^{(2)}_{P u}(
\mbox{\boldmath{$q$}}_1,
\mbox{\boldmath{$q$}}_2) &=& \sum_{\alpha} \left| T_{\alpha}^{ (2)}
\left(\widetilde{q}_1, \widetilde{q}_2\right)
\right|^2
 (2\pi)^4\delta^4\left(P_\alpha+\widetilde{q}_1+\widetilde{q}_2-P\right)
\nonumber\\
&=& \int d^4 x_1 d^4 x_2\,
g_P^{\left(2\right)}\left(x_1,\widetilde{q}_1;x_2, \widetilde{q}_2\right).
\nonumber
\end{eqnarray}
If we now demand that $g^{(2)}_P$ be normalized for on-shell emission
\begin{equation}\label{eq:norm}
\int \frac{d^3 q_1}{(2\pi)^3} \int \frac{d^3 q_2}{(2\pi)^3}
g^{(2)}_P(x_1,\widetilde{q}_1;x_2,\widetilde{q}_2) \equiv 1,
\end{equation}
we see that the normalization factor $W_{P u}$ introduced in
Eq.\ (\ref{eq:6}) is just the unsymmetrized total decay rate.
We thus find for this correlation function
\widetext
\begin{equation}\label{eq:10a}
C^{\text{l/u}}(\mbox{\boldmath{$q$}}_{\text{av}},
\mbox{\boldmath{$q$}}_{\text{rel}})
 = 1 + {\int d^4 x_1 d^4 x_2
g_P^{\left(2\right)}\left(x_1,q_{\text{av}};x_2, q_{\text{av}}\right)
\cos (q_{\text{rel}}(x_1-x_2))
\over
\int d^4 x_1 d^4 x_2
g_P^{\left(2\right)}\left(x_1,\widetilde{q}_1;x_2, \widetilde{q}_2\right)
 }.
\end{equation}
\narrowtext

To get an expression for the correlation function in terms of one-particle
quantities (c.f. Eq.\ (\ref{eq:Pratt})) the
two-particle Wigner function must be factorized into one-particle
functions.  This demands a number of assumptions.  In
particular, we require that:
\begin{itemize}
\item[(A1)]
the unsymmetrized T-matrix factorizes
\begin{equation}\label{eq:9}
T^{\left(2\right)}_{\left(\alpha_1|\alpha_2|\alpha_3\right)}
\left(q_1,q_2\right)
=
T^{\left(1\right)}_{\alpha_1}\left(q_1\right)
T^{\left(1\right)}_{\alpha_2}\left(q_2\right)
T^{\text{bath}}_{\alpha_3}.
\end{equation}
This condition can be fulfilled if the production process is dominated
by tree graphs.  However, this is not enough to factorize
Eq.\ (\ref{eq:6}), because the $\delta$-function still couples the $\alpha_1$
and $\alpha_2$ subsystems.  So we also assume that:
\item[(A2)]
the summation over the $\alpha_3$ degrees of freedom serves as a
heat bath that effectively decouples the other two subsystems.
\end{itemize}
Under these two conditions the two-particle Wigner function
no longer depends explicitly on the energy and momentum of the
initial state but depends on the statistical distribution of
energy provided by the integration over $\alpha_3$.  In effect,
a microcanonical ensemble is replaced by a canonical ensemble
characterized by a temperature in the center-of-mass frame
$$  \sum_{\alpha_3} (2\pi)^4 \delta^4 (P_\alpha +\widetilde{q}_1
+\widetilde{q}_2 -P)
\sim  \exp ((E-P_1^0-P_2^0-\omega_1-\omega_2)/T). $$
We defined $P_i^0$ to be the energy for the degrees of freedom $\alpha_i$
and $\omega_i \equiv \omega\left(\mbox{\boldmath{$q$}}_i\right)
 \equiv \sqrt{\mbox{\boldmath{$q$}}_i^2 +
 m_{\pi}^2}$ is the energy of a meson with momentum
 $\mbox{\boldmath{$q$}}_i$.
 We label the corresponding Wigner functions $g_T$, i.e.,
\widetext
\begin{equation}\label{eq:9a}
  g^{(1)}_T(x_1,q_1) =\frac{1}{W_{T u}}
 \sum_{\alpha_1} \int {d^4 p_1 \over (2\pi)^4}
e^{-ip_1x_1-\left(P_1^0+\omega_1\right)/T}
 T^{(1)}_{\alpha_1}(q_1
+p_1/2)T^{(1)*}_{\alpha_1}(q_1-p_1/2) .
\end{equation}
\narrowtext
Then under (A1) and (A2) the two-particle
Wigner function factorizes as
\begin{equation}\label{eq:10}
g_{\text{T}}^{\left(2\right)}\left(x_1,q_1;x_2,q_1\right)
 =
 g_{\text{T}}^{\left(1\right)}\left(x_1,q_1\right)
 g_{\text{T}}^{\left(1\right)}\left(x_2,q_2\right) .
\end{equation}

The requirements
(A1) and (A2) correspond to the usually cited assumption that
the meson field is {\it chaotic}, i.e., the mesons are created independently.
Under these conditions we can now identify the like-to-unlike correlation
function, Eq.\ (\ref{eq:l/u}),  with Pratt's expression,
Eq.\ (\ref{eq:Pratt}).
However, the two conditions above do not suffice to rewrite the
doubles-to-singles correlation function, Eq.\ (\ref{eq:d/s}), in the same way.
In addition we have to require that:

\begin{itemize}
\item[(A3)] the one-particle decay rate can be calculated
neglecting the symmetrization of the amplitude.
\end{itemize}
As this assumption is independent of (A1) and (A2) we can state it
in terms of the microcanonically calculated function as defined below
Eq.\ (\ref{eq:3})
\widetext
\begin{equation}\label{eq:11}
W^{\left(1\right)}_P \left(
\mbox{\boldmath{$q$}}_1\right)\approx
W^{\left(1\right)}_{P u} \left(
\mbox{\boldmath{$q$}}_1\right) =
\int \frac{d^3 q_2}{\left(2\pi\right)^3} \sum_{\alpha}
\left|T^{\left(2\right)}_{\alpha}\left(\widetilde{q}_1,
\widetilde{q}_2\right)\right|^2
\left(2\pi\right)^4
\delta^{(4)}\left(P_{\alpha} + \widetilde{q}_1+
\widetilde{q}_2 -P\right)
\end{equation}
\narrowtext
A further consequence of  (A3) is that
the total decay rate $W_P$ can be determined
neglecting symmetrization as well, i.e, $W_P \approx W_{P u}$.
It is usually assumed that this condition is fulfilled if typical
momenta of the participating particles are large compared to
the inverse size of the system, i.e., that the emission processes are
sufficiently localized in the emission volume. This condition, however,
can be misleading. We will come back to this point in Sec.\ \ref{sec:singles}.

Given assumption (A3) in addition to (A1) and (A2) the
doubles-to-singles correlation function, Eq.\ (\ref{eq:d/s}), can be written
using one-particle Wigner functions as
in Pratt's expression, Eq.\ (\ref{eq:Pratt}).

\section{A solvable model}\label{sec:model}
In this section we will describe a simple model that we use
to test how well the assumptions (A1)-(A3)
can be fulfilled
in a quantum mechanical calculation of meson production by
resonance decay. The model is based on a picture of a small source
that emits heavy particles.  These undergo
a cascade of decays
to reach the final state.  This picture is a plausible one for
the decay of the concentrated high energy zone produced by
nucleon-antinucleon annihilation or by $e^+$-$e^-$ annihilation
at high energy.  The mesonic source would have a spatial
extension depending on the initial size of the state and the
distance the resonances propagate.  In heavy ion collisions,
an additional mechanism extending the source is the rescattering
of mesons in the later stages of the evolution.  In our
model, we will permit only the resonance propagation to
contribute to the classical source size.  The model is thus
quite unrealistic for the heavy ion studies, but the criteria
we develop have a more general validity.

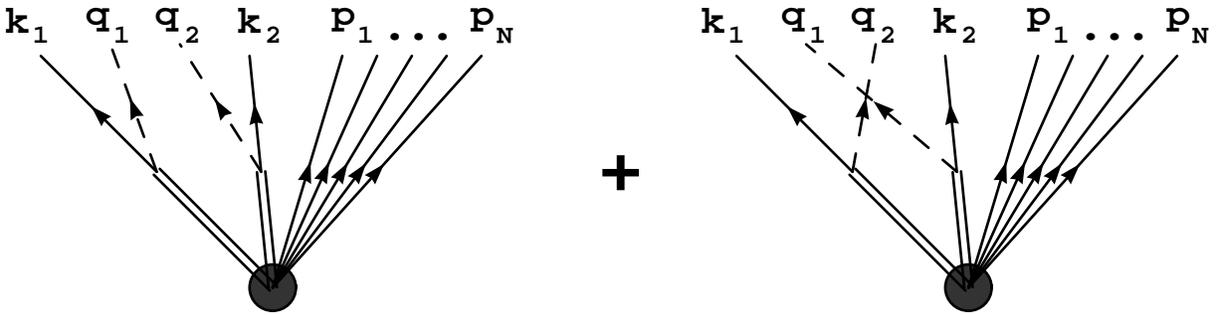
\begin{figure}
\caption{Graphical representation of the symmetrized
 amplitude for the production of two mesons in our model.
\label{fig:1}
}
\end{figure}

We will further simplify the resonance cascade process to the
Feynman graphs shown in Fig.\ \ref{fig:1}.
Here an
excitation having definite energy and momentum
decays into two resonances and $N$ additional particles.
Each resonance propagates and emits a meson into the
final state.  Only the two mesons are observed; the other
particles will be integrated out in calculating the differential
decay rates.  Note that this graph satisfies the condition
(A1) with the three sets of quantum numbers given by the
momenta $k_1$, $k_2$, and the set $p_i$.

For computational convenience we assume that all unobserved
particles have the same mass $m_f$, and ignore effects of symmetry
among the unobserved particles. We further assume that we can
treat the heavy particles nonrelativistically. Their on-shell
energy is then given by $\epsilon_f\left(
\mbox{\boldmath{$k$}}\right)\equiv
\mbox{\boldmath{$k$}}^2/ 2 m_f$.

The intermediate resonances are assumed to be represented by
simple resonance energy denominators of the form
\begin{equation}\label{eq:22}
G_{r}\left(E,\mbox{\boldmath{$P$}}\right) \equiv
\frac{1}{E-\epsilon_{r}\left(
\mbox{\boldmath{$P$}}\right)-m_{\text{diff}}+
i \Gamma_{r}/2}
\end{equation}
where $\epsilon_{r}\left(
\mbox{\boldmath{$P$}}\right) \equiv
\mbox{\boldmath{$P$}}^2/2 m_{r}$,
and
$m_{\text{diff}}\equiv m_{r}-m_f$ is the difference between
resonance mass and the mass of the unobserved particle in the final
state.
The quantity $\Gamma_{r}$ is the width of the resonance.
In
principle, the coupling constants should be such that the decay
rates are consistent with the assumed widths, but since the
absolute decay rates play no role we shall  ignore this
requirement and choose the coupling constant being one.

The couplings at the vertices
will be taken as point-like, since we are ignoring finite sizes of
the initial source and of the particles.  The meson-production
vertices contribute factors of $1/\sqrt{2\omega(\mbox{\boldmath{$q$}})}$
in the perturbation theory.  For the numerical calculations, we choose
parameters to correspond to the $\pi$ meson, emitted from the
$\Delta$ resonance of the nucleon.
Accordingly,  $m_\pi = 139 \text{ MeV}, m_f = 939 \text{ MeV}, m_{r} =
1232 \text{ MeV}$,
 and $\Gamma_{r}\equiv\Gamma_{\Delta} = 115 \text{ MeV}$.
However, note that our assumption of constant form factors
corresponds to an s-wave rather than the physical p-wave
character of the $\Delta$ resonance.

Finally, we will consider only one spatial dimension in the
remainder of this work.  We will switch to a notation with the
space or momentum variables denoted with a Roman letter and the
time or energy variable explicitly written out rather than
subsumed in a relativistic notation.

The differential decay rate associated with Fig.\ \ref{fig:1}
 is given by the following expression
\begin{eqnarray} \label{eq:model}
&&W_P^{\left(2\right)}\left(q_1,q_2\right)\nonumber\\
&&  =
\frac{1}{2}
\frac{1}{2\omega\left(q_1\right)2\omega\left(q_2\right)}
\int \frac{dk_1}{2\pi} \frac{dk_2}{2\pi}  \prod_{j=1}^N \frac{dp_j}{2\pi}
\left(2\pi\right)^2 \delta(E-E_1-E_2-\sum_{i=1}^N
\epsilon_f(p_i))\delta(P-P_1-P_2 - \sum_{i=1}^N p_i)
\nonumber\\
& & \times \left|\vphantom{\int}
 G_{r}(E_1 ,P_1)
 G_{r}(E_2,P_2)\right.\nonumber\\
& & \left.\vphantom{\int} +
 G_{r}(E_1+\omega(q_2)-\omega(q_1),P_1+q_2-q_1)
 G_{r}(E_2+\omega(q_1)-\omega(q_2),P_2+q_1-q_2)
\right|^2
\end{eqnarray}
For notational convenience we have defined
quantities
\begin{equation}
E_i \equiv \epsilon_f(k_i)+\omega(q_i)
\end{equation}
 and
\begin{equation}
P_i \equiv q_i+k_i
\end{equation}
 corresponding to the resonance energies and
momenta, respectively.
In the next section we will examine how this expression
approaches $W_T$ as the number of particles $N$ gets large.

\section{Thermal limit}\label{sec:thermlim}

Our first application of the model will be to see how the thermal
limit is approached as the number of spectator particles increases.
We first define the spectator phase space volume by integrating
the delta-functions in Eq.\ (\ref{eq:model}) over the spectator
momenta $p_j$,
 \begin{equation}\label{eq:18}
\Omega^{\left(N\right)}\left(E,P\right) \equiv
\int \frac{ dp_1 \ldots dp_N}{\left(2\pi\right)^N}
\left(2\pi\right)^2
\delta\left( E-\sum_{i=1}^{N} \epsilon_f\left(p_i\right)\right)
\delta\left(P-\sum_{i=1}^{N} p_i\right).
\end{equation}
This may be evaluated in closed form to obtain\footnote{The
$N$-fold integration can only be done analytically for
nonrelativistic particles.}
\begin{equation}\label{eq:20}
\Omega^{\left(N\right)}\left(E,P\right) =
 K
\left(E  - \frac{P^2}{2 N m_f}
\right)^{\frac{N-3}{2}}.
\end{equation}
Here K is a constant that we calculated but, as the expression is
complicated and will drop out later on, we do not show it here.
Inserting this in Eq.\ (\ref{eq:model}), the two-particle
probability distribution can be evaluated as a double
integral.  Before presenting the numerical results it is
useful to see how the thermal limit arises.  When $N$ is
large we may write the needed phase space function as
\widetext
$$
\Omega^{(N)}(E-E_1-E_2,P-P_1-P_2)\approx \Omega^{(N)}(E,P)
\exp (-N(E_1+E_2-(P_1+P_2)\frac{P}{m_f N})/2 E^*) ,
$$
\narrowtext
where $E^* = E - P^2/2 m_f N$ is the center-of-mass energy.
{}From here on  we shall consider momenta in the frame where
$P\equiv 0$.
  The effective
temperature associated with the phase space is then
$T= 2E/N$, as expected from the equipartition theorem for
$N$ degrees of freedom.

We now compare the doubles-to-singles correlation function
evaluated explicitly from the finite-$N$ two-particle
probability function Eq.\ (\ref{eq:model}), and the
corresponding thermal limit,
\widetext
\begin{eqnarray} \label{eq:thermal}
& & W_T^{\left(2\right)}\left(q_1,q_2\right)\nonumber \\
 && =  \frac{\Omega^{(N)}(E,P=0)}{2}
\frac{1}{2 \omega\left(q_1\right) 2 \omega\left(q_2\right)}
\int \frac{dk_1}{2\pi} \frac{dk_2}{2\pi}
 \exp\left(-\left(E_1+E_2\right)/T\right)
\nonumber \\
 & & \times \left| \vphantom{\int}
 G_{r}(E_1 ,P_1)
 G_{r}(E_2,P_2)
\right. \nonumber \\
& + & \left.\vphantom{\int}
 G_{r}(E_1+\omega(q_2)-\omega(q_1),P_1+q_2-q_1)
 G_{r}(E_2+\omega(q_1)-\omega(q_2),P_2+q_1-q_2)
\right|^2 .
\end{eqnarray}
\narrowtext

Fig.\ \ref{fig:2}
shows typical results for $C^{\text{d/s}}$
calculated with Eqs.\ (\ref{eq:model}) and (\ref{eq:thermal}).
The total energy was chosen  to be $E =  1.5\,m_\pi\,N$
corresponding to a temperature of
$T = 3\,m_\pi$, which is the temperature used in
Eq.\ (\ref{eq:thermal}).
 The correlation function is shown as a function of the
momentum difference $q_{\text{rel}}=q_1-q_2$ for fixed average momentum
$q_{\text{av}} = 0$. It may be seen that the finite-$N$
correlation function approaches the result of the thermal limit.
So we conclude that as few as 50 participating particles justify
the assumption that the correlation function can be
calculated using a canonical ensemble instead of a microcanonical
one.  We expect that in three dimensions the same heat bath would
require of the order of 20 particles.

\section{The singles distortion}\label{sec:singles}

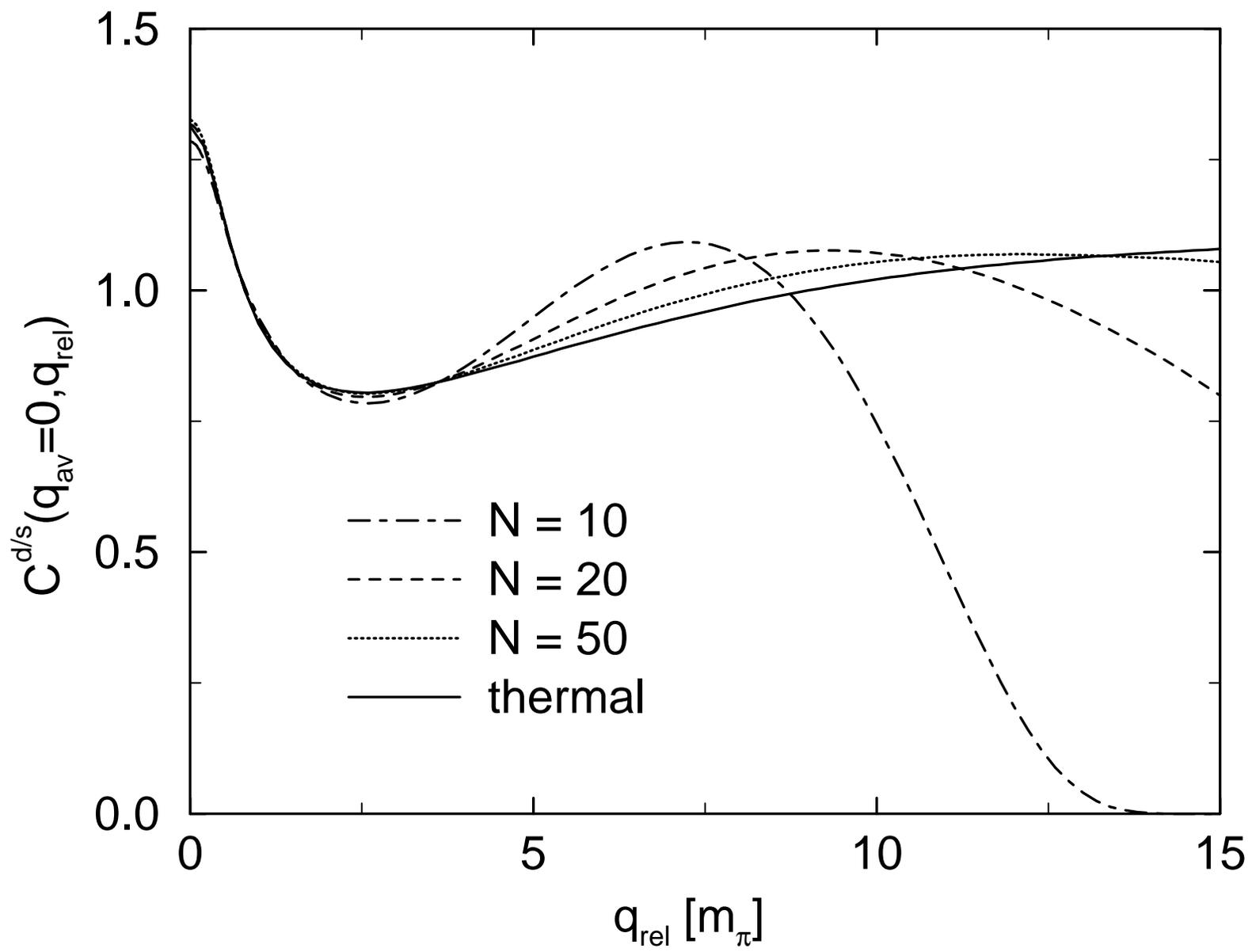
\begin{figure}
\caption{Correlation function $C^{\text{d/s}}$ for different numbers
of particles and in the thermal limit. The  energy
is $ E = 1.5 N m_\pi $
 which corresponds in the thermal limit to a temperature of
$T = 3\,m_\pi$. The average momentum of the meson pair is zero for all
cases.
\label{fig:2}
}
\end{figure}

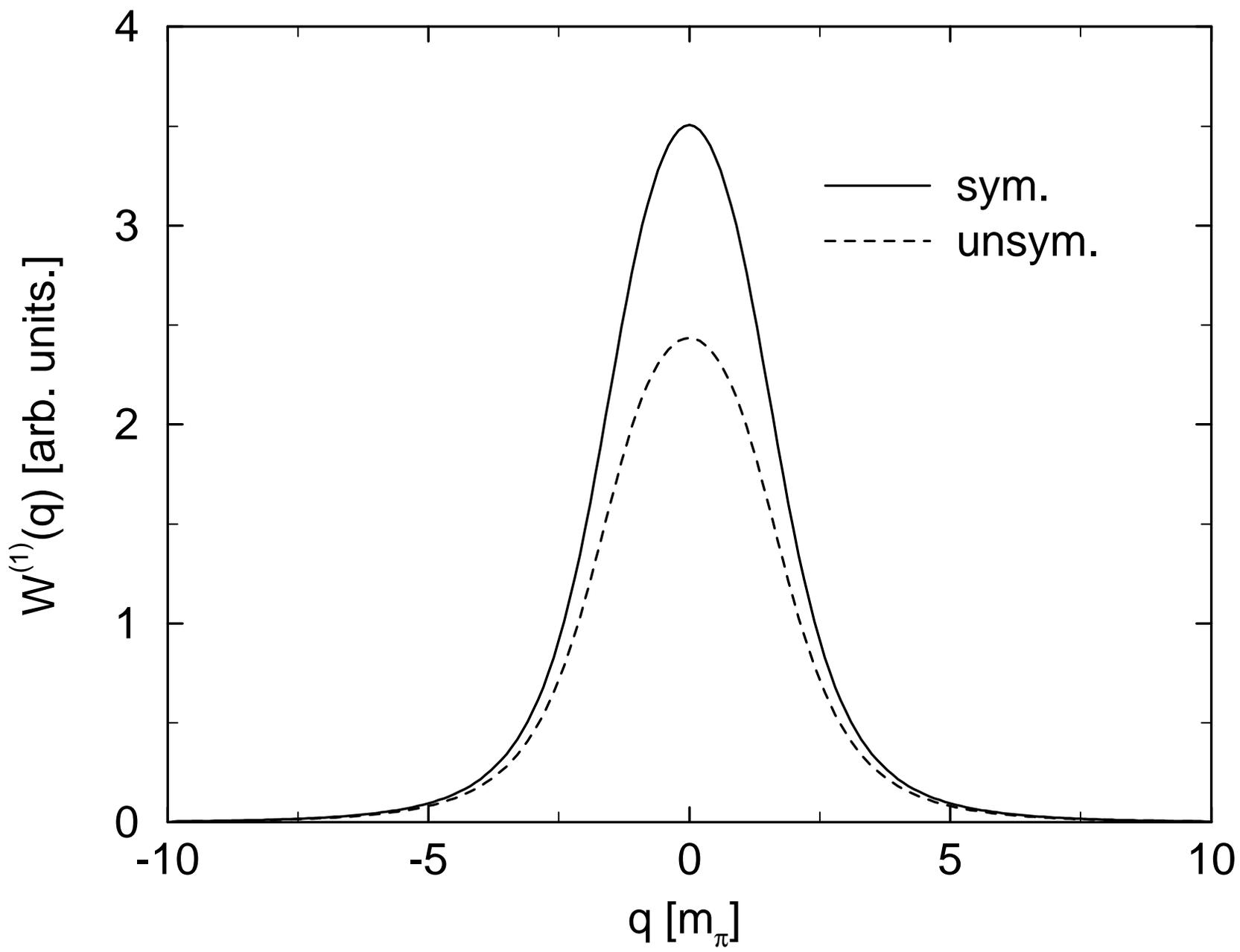
\begin{figure}
\caption{
Single particle decay rate $W^{(1)}(q)$ calculated with the
symmetrized (full line) and unsymmetrized (dashed line)
two-particle amplitude, for $T = 3 m_{\pi}$ and $\Gamma_r = \Gamma_\Delta$.
\label{fig:3}
}
\end{figure}

It is immediately apparent from Fig.\ \ref{fig:2} that the correlation
function looks quite different from the expected form, which
should start at 2 at zero $q_{\text{rel}}$ and fall monotonically to
1 at large $q_{\text{rel}}$.  We can trace this behavior
back to a violation
of condition (A3). In Fig.\ \ref{fig:3} we show $W_T^{\left(1\right)}$
(full line)
calculated by integrating over the symmetrized $W^{(2)}_T$
as compared to an integral over the unsymmetrized $W^{(2)}_{T u}$
(dashed line) (cf.\ Eqs.\ (\ref{eq:3}) and (\ref{eq:11})).
The two curves clearly deviate strongly for small values of $q$.
It is obvious that also the total particle production probability
can not be calculated neglecting symmetrization. These facts account
for the {\it wrong} asymptotic values of the correlation function.
To make things worse the correlation function has a pronounced dip
for intermediate values of $q_{\text{rel}}$. That means that even
the introduction of an overall normalization factor and a
coherence parameter $\lambda$ does not help to extract
the desired information about the source from $C^{\text{d/s}}$.

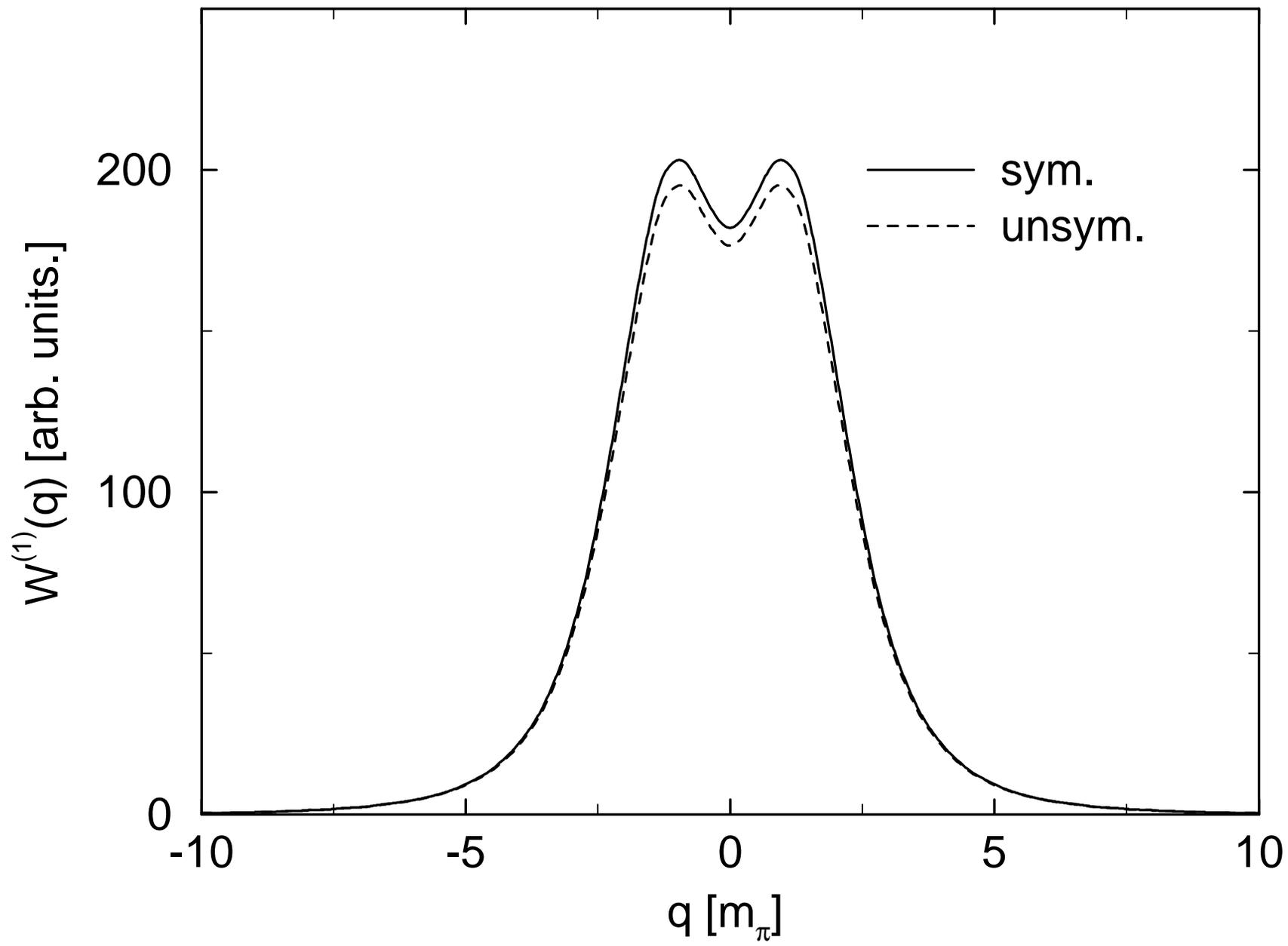
\begin{figure}
\caption{
Same as Fig.\ \protect{\ref{fig:3}}, but for a small resonance width
$\Gamma_r = \Gamma_\Delta/10$.
\label{fig:5}
}
\end{figure}

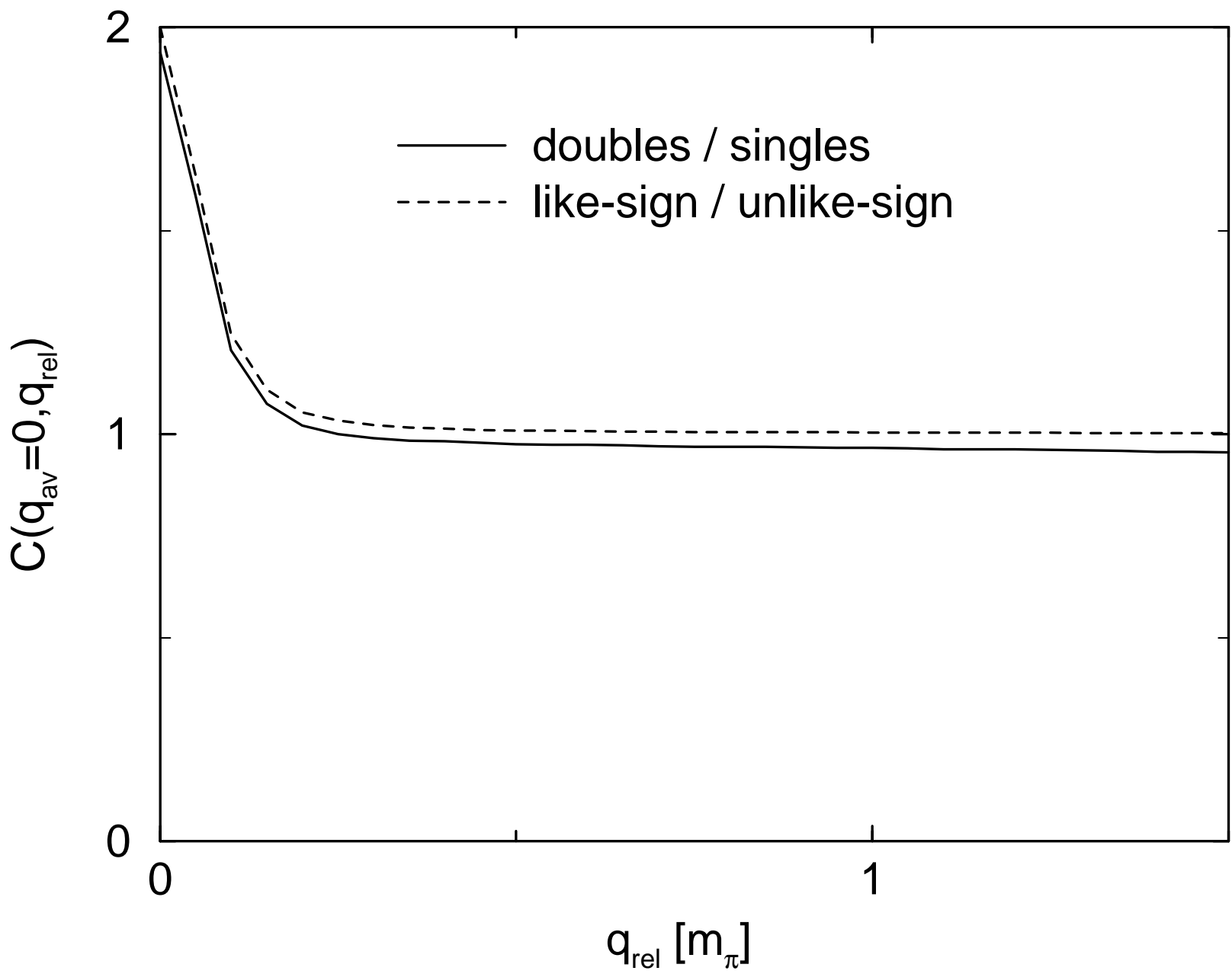
\begin{figure}
\caption{
The correlation functions $C^{\text{d/s}}$ (full line) and $C^{\text{l/u}}$
(dashed line) for a resonance width $\Gamma_r = \Gamma_\Delta /10$ and
$T = 3 m_\pi$.
\label{fig:6}
}
\end{figure}

The validity of (A3) can be determined empirically\cite{Za84}. The
single particle distribution is evaluated by integrating over one
of the two momenta in the two-particle distribution function Eq.\ (\ref{eq:8})
or (\ref{eq:thermal}). The contribution of the interference term
is of the order of the width of the correlation function $\Delta q_c$
divided by a typical momentum difference\\
 $<(q_1-q_2)^2>^{1/2}$. This
should be small for (A3) to hold. From Fig.\ \ref{fig:2} we have
$\Delta q_c \approx 3/4 m_\pi$ and
Fig.\ \ref{fig:3} gives $<(q_1-q_2)^2>^{1/2} \approx 3 m_\pi$,
yielding
\begin{equation}\label{eq:x}
\frac{\Delta q_c}{<(q_1-q_2)^2>^{1/2}} \approx 25 \% .
\end{equation}
This is intolerably large to ignore in the HBT analysis. One might
anticipate that in three dimensions the relevant expression would
be the cube of Eq.\ (\ref{eq:x}), which would make (A3) a rather
good assumption. However, analysis of experimental data shows
distortions of the order of 10 \% \cite{Hu93}. Also in a Fritiof Monte Carlo
simulation of NA35 data \cite{KS92} an influence of the symmetrization
of the singles spectrum of 10-20 \% was found.
Coming back to our model, we find that
by decreasing
the width of the resonance down to $\Gamma_{r}
 = \Gamma_\Delta/10$ we actually
decrease $\Delta q_c$
 so much that (A3) becomes valid. In Fig.\ \ref{fig:5} we show
the single particle decay rate again for the symmetrized (full line) and
unsymmetrized (dashed line) two particle amplitude. Compared to
Fig.\ \ref{fig:3} the deviation of the two curves is very small.
Indeed the  calculation of the doubles to singles correlation function
yields a satisfactory result, shown in Fig.\ \ref{fig:6} (full line).
The small deviation from two at low $q_{\text{rel}}$ and one for high
$q_{\text{rel}}$ is due to the residual small difference
between the single particle decay rate in the symmetrized and the
unsymmetrized case respectively as well as numerical inaccuracy in evaluating
the two-particle decay rate for such a small resonance width.
In Fig.\ \ref{fig:6} we show also the like-to-unlike correlation
function $C^{\text{l/u}}$ for the same parameters (dashed line).
As expected for
the case of a small singles distortion we find a good agreement of
the two correlation functions.

\section{Source sizes}\label{sec:size}

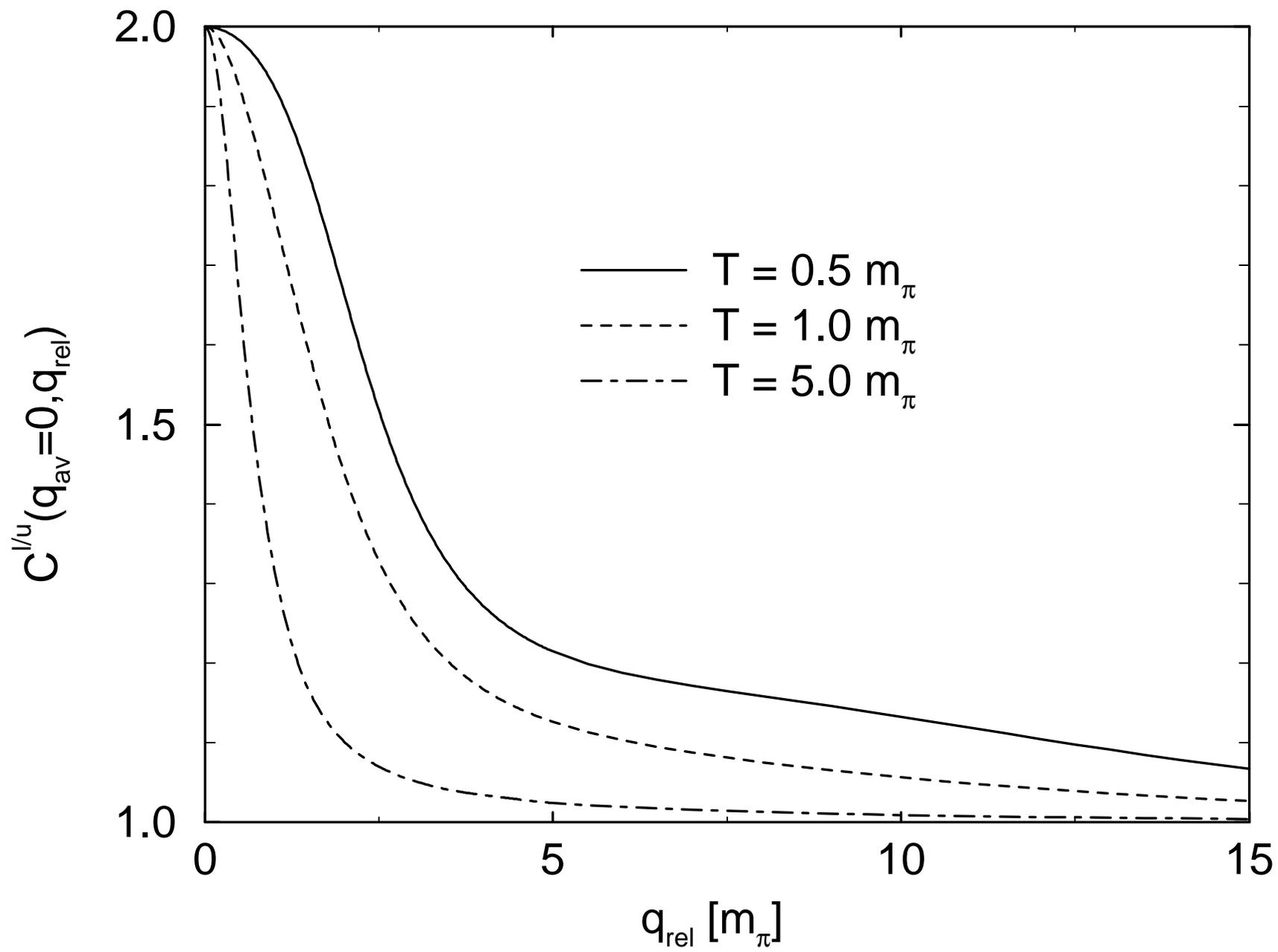
\begin{figure}
\caption{
The correlation function $C^{\text{l/u}}$
 for different temperatures $T = 0.5 m_\pi$ (full line),
$ 1.0 m_\pi$ (dashed line), and $5.0 m_\pi$ (dot-dashed line).
The width of the resonance is $\Gamma_r = \Gamma_{\Delta}$.
\label{fig:7}
}
\end{figure}

In the remainder of this work we shall concentrate on the measure
$C^{\text{l/u}}$ to avoid the problem of the distortion in the singles
spectrum.  In Fig.\ \ref{fig:7}
 we show the thermal model $C^{\text{l/u}}$ for parameters
corresponding to the physical $\Delta$ resonance and for several
values of the temperature.  The correlation function is
seen to be well-behaved, and one can try to extract source size
parameters from the width of the peak.  Notice that the
peak becomes the narrower as the temperature increases. This seems to indicate
that the emitting system is larger for higher temperatures. We will
discuss this point in detail in Sec.\ \ref{sec:res}.  We now
address the question of whether a classical source size can be extracted
from the peak width.

\subsection{Classical limit}\label{sec:classlim}
We first derive the classical limit of the one-particle
Wigner function for our model.
The Wigner transform of the quantum
amplitude has the form
\widetext
\begin{eqnarray} \label{eq:29}
&& g_{\text{T}}^{\left(1\right)}
\left(t,x,\omega_{\text{av}},q_{\text{av}}\right)\nonumber \\
&& =  \int \frac{dk}{2\pi}
 \text{e}^{-\left(\omega_{\text{av}}+\epsilon_f\left(k\right)
\right)/T}
 \int \frac{d\omega_{\text{rel}}}{2\pi} \frac{dq_{\text{rel}}}{2\pi}
\text{e}^{-i\left(\omega_{\text{rel}} t - q_{\text{rel}}x\right)}
\left(
2 \omega\left(q_1\right)
2 \omega\left(q_2\right)
\right)^{-1/2}
\nonumber \\
& & \times G_{r}\left(\epsilon_f\left(k\right)+
\omega_{\text{av}}+\omega_{\text{rel}}/2,
k+q_1\right)
G^*_{r}\left(\epsilon_f\left(k\right)+
\omega_{\text{av}}-\omega_{\text{rel}}/2,
k+q_2\right).
\end{eqnarray}
\narrowtext

The integration over $\omega_{\text{rel}}$ can be performed
by an ordinary contour integration.
The contribution from the two poles is
\begin{eqnarray} \label{eq:30}
& &g_{\text{T}}^{\left(1\right)}
\left(t,x,\omega_{\text{av}},q_{\text{av}}\right)\nonumber\\
& & =  \Theta\left(t\right)
\int \frac{dk}{2\pi}
 \text{e}^{-\left(\omega_{\text{av}}+\epsilon_f\left(k\right)
\right)/T}
 \int \frac{dq_{\text{rel}}}{2\pi}
\text{e}^{i q_{\text{rel}}x }
\left(
2 \omega\left(q_1\right)
2 \omega\left(q_2\right)
\right)^{-1/2}
\nonumber \\
& &
\times \frac{2 i \text{e}^{-\Gamma t}}{E_+ + E_-}
\left(\exp\left(-2 i E_+ t\right) - \exp\left(2 i E_- t\right)\right),
\end{eqnarray}

where
\begin{equation}\label{eq:31}
E_\pm = \epsilon_{r}\left(k+q_{\text{av}}\pm q_{\text{rel}}/2\right)
+ m_{\text{diff}} - \epsilon_f\left(k\right) - \omega_{\text{av}}.
\end{equation}
Up to this point the calculation is still exact, i.e., Eq.\ (\ref{eq:30})
still contains the full quantum mechanical information. We now
derive a classical source function by approximations on the above
expression.  We first expand $E_{\pm}$ to first order
in the momentum difference $q_{\text{rel}}$
\begin{equation}\label{eq:32}
E_{\pm}\approx \Delta E \pm
\left. \frac{d \epsilon_{r}\left(p\right)}{d p}
\right|_{p=q_{\text{av}}+k} q_{\text{rel}}/2 ,
\end{equation}
where
\begin{equation}\label{eq:33}
\Delta E = \epsilon_{r}\left(k+q_{\text{av}}\right) +
m_{\text{diff}} - \epsilon_f\left(k\right) -\omega_{\text{av}}.
\end{equation}
The derivative $d \epsilon_{r}\left(p\right)/d p$ is the
classical velocity $v_{r}\left(p\right)$ of the resonance.
Thus we may write
\widetext
\begin{equation}\label{eq:34}
g_{\text{T}}^{\left(1\right) \text{cl}}
\left(t,x,\omega_{\text{av}},q_{\text{av}}\right)
 = \frac{ \Theta\left(t\right)}{\omega\left(q_{\text{av}}\right)}
\text{e}^{-\Gamma_r t}\int \frac{dk}{2\pi}
 \text{e}^{-\left(\omega_{\text{av}}+\epsilon_f\left(k\right)
\right)/T}
 \int \frac{dq_{\text{rel}}}{2\pi}
\text{e}^{i \left(q_{\text{rel}}x
- q t v_{r}\left(k+q_{\text{av}}\right)\right) }
\frac{\sin\left(2 \Delta E t\right)}{\Delta E}.
\end{equation}
\narrowtext
The integral over $q_{\text{av}}$ gives a $\delta$-function
and the
last factor approaches the $\delta$-function
$2\pi \delta(\Delta E)$. Therefore
 the classical approximation amounts
to demanding that the resonance propagates on-shell.
Evaluating the $\delta$-functions and dropping the subscript
``av'' we obtain:
\widetext
\begin{equation}\label{eq:35}
g_{\text{T}}^{\left(1\right) \text{cl}}\left(t,x,\omega,q\right)
= \frac{\Theta\left(t\right)}{\omega\left(q\right)}
\text{e}^{-\Gamma_r t} \sum_{i=1,2}
\delta\left(x - v_{r}\left(k_f^i+q\right) t\right)
\text{e}^{-\left(k_f^i+q\right)^2/2 m_{r} T}
\frac{1}{\left|v_{r}\left(k_f^i+q\right)-v_f\left(k_f^i\right)
\right|} .
\end{equation}
\narrowtext
In this equation, $k_f^1$ and $k_f^2$ are the two roots of the condition
$\Delta E = 0$, which resulted from the evaluation of the
$k$-integral over the delta-function $\delta(\Delta E)$.
These roots are solutions of a quadratic equation and are given
by
\begin{equation}\label{eq:35a}
k_f^{1, 2}\equiv \frac{m_f}{m_{\text{diff}}} q
\pm \sqrt{\frac{m_f m_{r}}{m_{\text{diff}}^2} q^2 +
2 m_f m_{r}\left(1 - \frac{\omega}{m_{\text{diff}}}\right)}.
\end{equation}
The
k-integral gives the density-of-states factor at the end of
equation (\ref{eq:35}).
  Note that the various terms in this Wigner function
(except the density-of-states factors)
have obvious classical interpretations.  The initial $\Theta$
function specifies that the source starts at
$t=0$, and its exponential decay is given by the second factor.
The resonance propagates classically with connection between its
position, velocity, and the time given by the argument of the
delta-function.  The probability to make the resonance is given
by the Boltzmann factor following.
With this one-particle Wigner function in the classical approximation
we can evaluate the classical correlation function $C^{\text{cl}}$
by inserting $g_T^{(1) \text{cl}}$, Eq.\ (\ref{eq:35}), into Pratt's
expression, Eq.\ (\ref{eq:Pratt}).

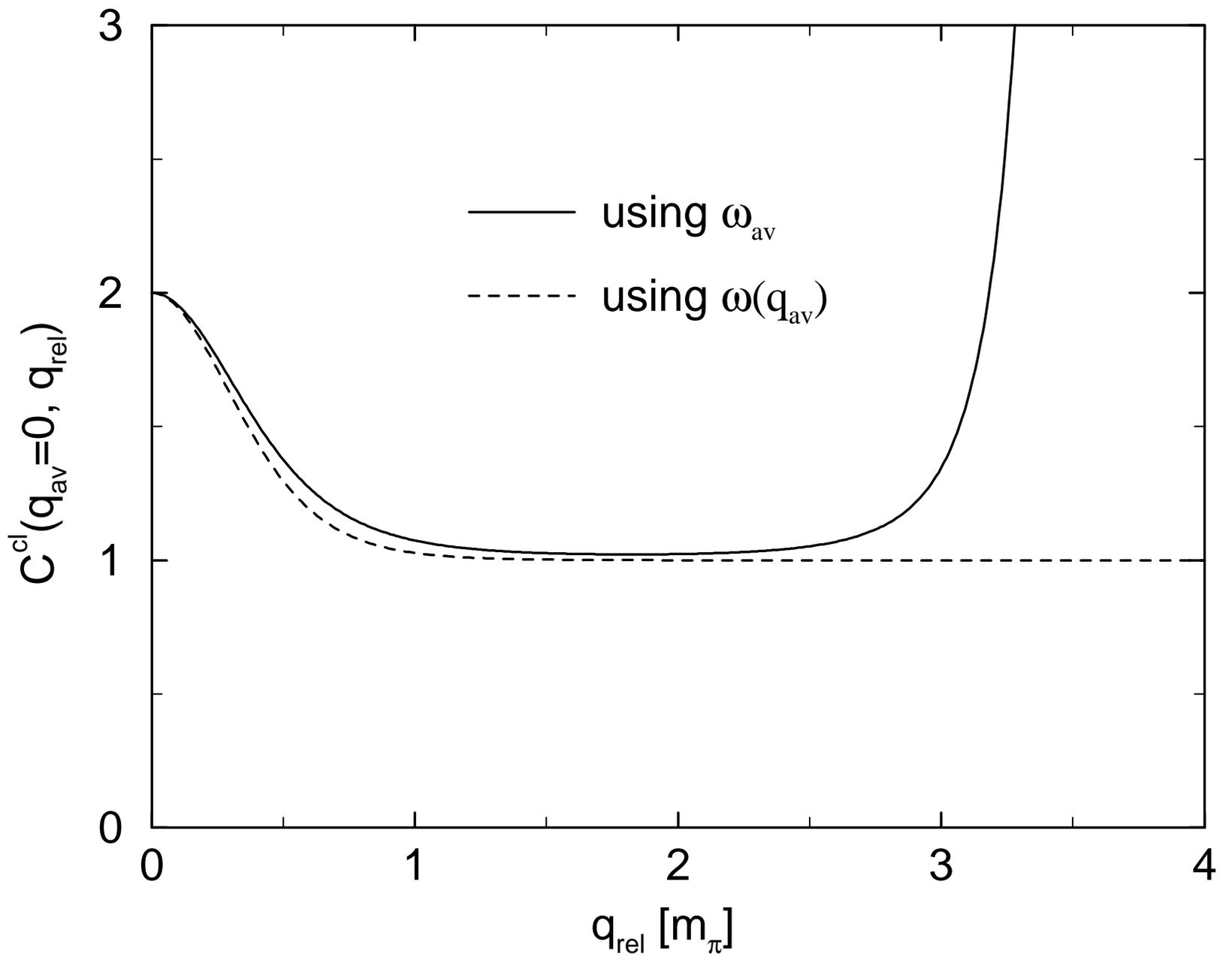
\begin{figure}
\caption{
Classical
correlation function with different treatments of the energy variable, for
$T = 1\,m_\pi$ and $\Gamma_r = \Gamma_\Delta$.
\label{fig:10}
}
\end{figure}

Note that
our classical $g^{cl}_T$ depends on the
variables $q$ and $\omega$ independently.
Fig.\ \ref{fig:10} shows the correlation as a function of
$q_{rel}$ with average momentum fixed at
$q_{\text{av}} = 0$ and temperature $T =  m_\pi$, indicated by the solid curve.
The correlation function behaves as expected for small
momentum differences. For large momenta, however, $C^{\text{cl}}$ develops
a singularity.
This can be traced back to the occurrence of $\omega_{av}$
in the argument of the
Wigner functions in the numerator of Eq.\ (\ref{eq:Pratt}).
Under sufficiently extreme conditions, this energy goes
off-shell so far that the classical condition $\Delta E= 0$ no
longer has a solution.  If $\omega_{av}$ is replaced by
an on-shell energy, namely by $\omega(q_{av})$, the problem
does not arise.  The correlation function calculated with
this on-shell prescription is shown as the dashed curve.
This function is very similar
to the original one for small $q_{\text{rel}}$. In contrast to
the singular behavior of the original correlation function
the dotted curve behaves smoothly also for high $q_{\text{rel}}$.
Since this problem arises at high momentum, and the source size
question depends on the low momentum behavior only, we can ignore
the differences for the remainder of this work.

\subsection{Contributions to the source size}
\label{sec:res}

A convenient way to characterize the
 information contained in the
correlation function is to take the second derivative with
respect to the momentum difference at $q_{\text{rel}} = 0$.
This gives an effective mean square source size defined
as
\begin{equation}\label{eq:rmseff}
<x^2>^{\text{eff}}\equiv -\frac{1}{2}
\left.
\frac{d^2 C}{  dq_{\text{rel}}^2} \right|_{q_{\text{rel}}=0} .
\end{equation}
We determine this from the single-particle Wigner source function
taking the derivative of Eq.\ (\ref{eq:Pratt}).  This yields
\widetext
\begin{eqnarray}\label{eq:37}
< x^2 >^{\text{eff}}
& = & \left[ \frac{1}{G} \int dt dx \left(x
-v_\pi\left(q_{\text{av}}\right) t
\right)^2 g - \left(\frac{1}{G}\int dt dx
 \left(x -v_\pi\left(q_{\text{av}}\right) t
\right) g \right)^2 -
\left(\frac{1}{2 G}\frac{dG}{dq_{\text{av}}}\right)^2
\right.
\nonumber \\
& & \left.
+\frac{1}{4 G}\frac{d^2 G}{d q_{\text{av}}^2}
-\frac{1}{4 G}\left.\frac{d^2\omega\left(q\right)}{d q^2}
\right|_{q=q_{\text{av}}}\frac{\partial}{\partial \omega}
\int dt dx \left. g\left(t,x,\omega,q_{\text{av}}\right)\right|_{
\omega = \omega (q_{av})}\right].
\end{eqnarray}
\narrowtext
In writing this we used the abbreviations $g = g^{\left(1\right)}\left(t,x,
\omega(q_{av}),q_{\text{av}}\right)$ and  $G = \int dt dx
g^{\left(1\right)}\left(t,x,\omega (q_{av}),q_{\text{av}}\right)$.
The quantity $v_{\pi}$ is the velocity of the pion $v_{\pi}(q) \equiv
d \omega(q)/d q$.
We will refer to the five contributions of the r.h.s. of Eq.\ (\ref{eq:37})
by $<x^2>^{I}, \ldots, <x^2>^{V}$ respectively.
$<x^2>^{I}$ and $<x^2>^{II}$ correspond
to the space-time distribution of the emission points of the mesons.
This is the information we would like to extract.
The other terms give corrections which are due to the energy and momentum
dependence of the source.

To simplify the discussion we concentrate on the behavior at $q_{\text{av}}=0$.
Then $<x^2>^{II}$ and $<x^2>^{III}$ in Eq.\ (\ref{eq:37})
vanish by symmetry.
The first term is easily evaluated with the classical source
$g_T^{(1) \text{cl}}$ (Eq.\ (\ref{eq:35})).  The result is
\begin{eqnarray}\label{eq:37a}
<x^2>_{\text{cl}}^I &=& \frac{1}{G} \int dt dx \left(x
-v_\pi\left(q_{\text{av}}\right) t
\right)^2 g_T^{(1) \text{cl}} \nonumber \\
&=& 2\left(k^0_f/m_r \over \Gamma_r \right)^2 ,
\end{eqnarray}
where $k_f^0$ is the classical momentum of the resonance required
to emit a meson at $q=0$.
It is given by (compare Eq.\ (\ref{eq:35a}))
\begin{equation}\label{eq:39}
k_f^0 \equiv \sqrt{2 m_f m_{r} \left(1 - \frac{m_\pi}{m_{\text{diff}}}
\right)}.
\end{equation}
The result Eq.\ (\ref{eq:37a}) is
of course exactly what one would expect for a source moving
with velocity $k^0_f/m_r$ and decaying at a rate $\Gamma_r$.

Note also that the source size is independent of the temperature. This
is counterintuitive at the first glance. However, in the classical
approximation the momentum of the resonance is kinematically
fixed by the momentum of the emitted pion.
Therefore  the distance the resonance travels
before it emits the pion is determined only by the kinematics
and lifetime.

\subsection{Comparison of classical and quantum results}\label{qvscl}

We now extract the source size from the quantum mechanical
thermal source.
The correlation function at $q_{\text{av}} = 0 $ has the form
\begin{equation}\label{eq:40}
C^{\text{l/u}}\left(q_{\text{av}}=0,q_{\text{rel}}\right)
= 1 + \frac{\left| {\cal I}_T
\left(q_{\text{rel}},-q_{\text{rel}}\right)\right|^2}
{
{\cal I}_T\left(q_{\text{rel}},q_{\text{rel}}\right)
{\cal I}_T\left(-q_{\text{rel}},-q_{\text{rel}}\right)
}
\end{equation}
where
\widetext
\begin{eqnarray}
{\cal I}_T\left(q_1,q_2\right) &=& \int dk\ \text{e}^{-\left(\epsilon_f(k)
+\omega(q_1/2)\right)/T}\nonumber\\
&&\times
 G_r(\epsilon_f(k)+\omega(q_1/2),k+q_1/2)\ G_r^*(\epsilon_f(k)+\omega(q_2/2),
k+q_2/2) .
\end{eqnarray}
\narrowtext
It is straightforward to evaluate ${\cal I}$ by contour integration.
The result in the $T\rightarrow \infty$ limit is particularly
simple, yielding for the correlation function
\begin{equation}\label{eq:41}
C^{\text{l/u}}_{\infty}\left(q_{\text{av}}=0,q_{\text{rel}}\right)
= 1 + \frac{1}{\left(1 + \left(
 q_{\text{rel}}/q_0\right)^2
\right)^2} ,
\end{equation}
where we have defined
\begin{equation}\label{eq:42}
q_0
 \equiv  \text{Im} \sqrt{\frac{m_{r}}{m_f}
q_{\text{rel}}^2 + 8 \frac{m_{\text{diff}} m_{r}}{m_f}
\left(m_{\text{diff}} - \omega\left(q_{\text{rel}}/2\right)
- i \Gamma_{r}/2\right)} .
\end{equation}
The source size may be evaluated from Eq.\ (\ref{eq:rmseff}).
In the limit $\Gamma_r \ll m_{\text{diff}} - m_{\pi}$,
 it reduces exactly to the
classical formula, Eq.\ (\ref{eq:37a}),
\begin{equation}\label{eq:45}
<x^2>^{\text{eff}}_q
\approx  <x^2>_{\text{cl}}^{I} .
\end{equation}

At finite temperature, the expression for $C^{\text{l/u}}$ is rather
unwieldy and we quote only the source size, obtained either directly
from Eq.\ (\ref{eq:rmseff}) or with Eq.\ (\ref{eq:37}) and the quantum Wigner
function Eq.\ (\ref{eq:29}). The quantum mean square radius
$<x^2>^{\text{eff}}_q$
is found to be
\begin{equation}\label{eq:qmsr:qav}
<x^2>^{\text{eff}}_q = \left<\left<\frac{\left(v_r-v_{\pi}\right)^2}{
\left(\Delta E\right)^2 + \left(\Gamma_r/2\right)^2} \right>\right>
-\left| \left<\left<\frac{v_r -v_{\pi}}{\Delta E + i \Gamma_r /2}\right>\right>
\right|^2.
\end{equation}
Here we used
$\Delta E =
\epsilon_r (k+q_{\text{av}})
 + m_{\text{diff}} - \epsilon_f (k) - \omega(q_{\text{av}})$ and
the thermal average $\left<\left< f \right>\right>$
 weighted by the resonance distribution is defined
as
\begin{equation}\label{eq:aver}
\left< \left<\ f\ \right>\right> \equiv \
\frac{\int dk\ \text{e}^{-\epsilon_f (k)/T}
 f \left(
\left(\Delta E\right)^2 + \left(\Gamma_r/2\right)^2\right)^{-1}}{
\int dk\ \text{e}^{-\epsilon_f (k)/T}
\left(\left(\Delta E\right)^2 + \left(\Gamma_r/2\right)^2\right)^{-1}} .
\end{equation}

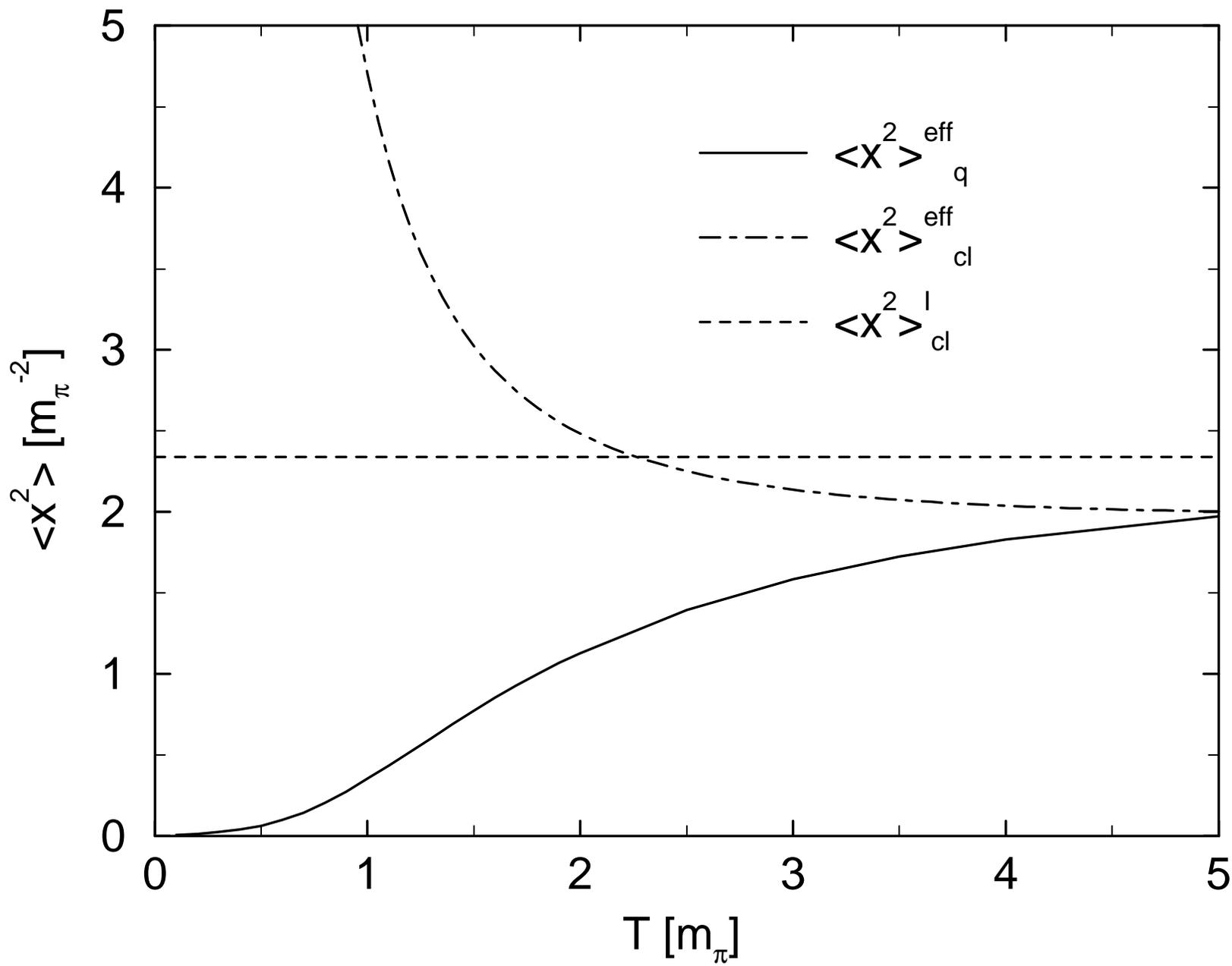
\begin{figure}
\caption{
The mean square source size as a function of the
temperature in various treatments.
The horizontal dashed line is the classical size $<x^2>_{\text{cl}}^{I}$,
 Eq.\ (\protect{\ref{eq:37a}}).
The dot-dashed curve includes corrections to the size of the classical
source from Eq.\ (\protect{\ref{eq:37}}). The solid line gives the result
from the quantum mechanical calculation, Eq.\ (\protect{\ref{eq:qmsr:qav}}).
 The width is $\Gamma_{r} = \Gamma_{\Delta}$.
\label{fig:11}
}
\end{figure}

In Fig.\ \ref{fig:11} we show the mean square source size for a width of
the resonance $\Gamma_r = \Gamma_{\Delta}$ and $q_{\text{av}}=0$.
Note that for vanishing average momentum only the first term
on the r.h.s of Eq.\ (\ref{eq:qmsr:qav}) contributes.
 The quantum mean square
radius $<x^2>^{\text{eff}}_q$ is shown as a function of temperature,
indicated by the solid line. This is compared to the
temperature-independent classical size
$\left<x^2\right>_{\text{cl}}^{I}$
indicated by  the dashed line.
 It might be hoped that the
additional terms in Eq.\ (\ref{eq:37}) would improve the classical description,
but this is not the case. The full classical source size
$<x^2>^{\text{eff}}_{\text{cl}}$
, shown as the dot-dashed line,
in fact deviates in the opposite direction from the quantum result. The
disagreement is serious, since we are interested in temperatures of the order
of $T = m_{\pi}$ and lower, for physical sources.

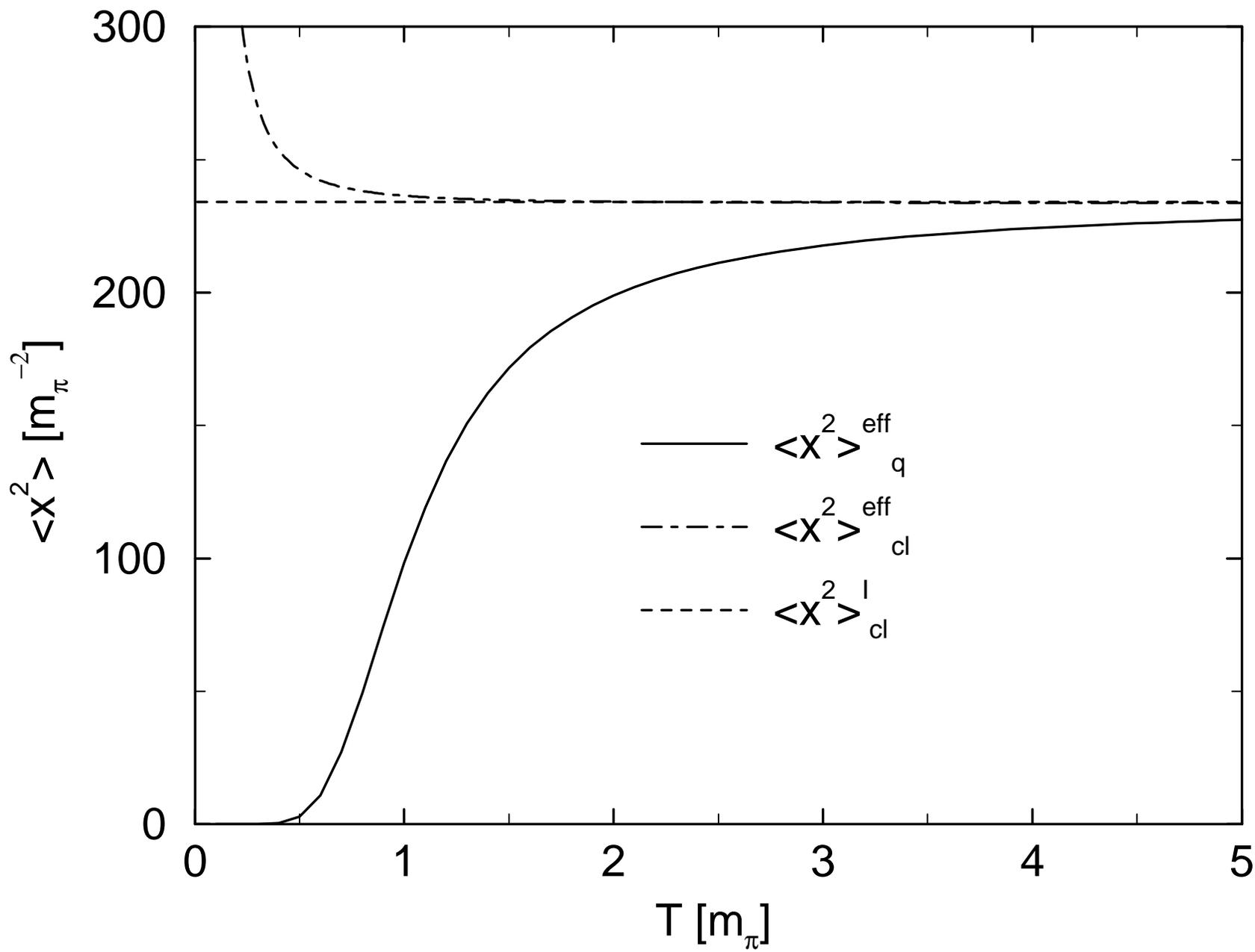
\begin{figure}
\caption{
Same as Fig.\ \protect{\ref{fig:11}} but for the reduced width
$\Gamma_{r} = \Gamma_\Delta/10$.
\label{fig:12}
}
\end{figure}

Clearly, the off-shell propagation is important at physically realizable
temperatures. Even for widths that might seem small, the exponential
factor in Eq.\ (\ref{eq:aver}) emphasizes states with low energy in
Eq.\ (\ref{eq:qmsr:qav}).
This is illustrated in Fig.\ \ref{fig:12}, showing the source size for
a very small width, $\Gamma_r = \Gamma_\Delta/10$. We see that even
in this case the classical source is unreliable for temperatures
below $2 m_\pi$.
The problem is severe because in the classical calculation
a  zero-momentum pion is not produced  from a low momentum resonance as
is favored in the quantum mechanical calculation.
However, if we choose conditions such that the
resonance region is not suppressed by the Boltzmann factor,
the classical source is accurate down to much lower temperatures.

Even though  our model calculation shows that quantum effects are important
for the interferometry among mesons produced by resonances, our result
 Eq.\ (\ref{eq:qmsr:qav}) suggests a way to improve classical modeling.
When particles propagate far off-shell, their lifetime is controlled
by $(\Delta E)^{-1}$, according to the uncertainty principle,
rather than by $\Gamma_r^{-1}$. Eq. (\ref{eq:qmsr:qav}) could be recovered
from a classical simulation if in Eq.\ (\ref{eq:37a}) the resonance
lifetime $\Gamma_r^{-1}$
 is replaced by an effective lifetime
 $ \sim ((\Gamma_r/2)^2 + (\Delta E)^2)^{-1/2}$,
with a suitable averaging over an ensemble of resonances with different
$\Delta E$'s.

\section{Summary}\label{sec:summary}
We have studied a resonance decay model to test the
statistical assumptions behind the HBT analysis of meson
correlations.  One set of assumptions gives rise to a reduction
of the unsymmetrized two-particle source function to a
product of two single-particle thermal sources.  We find that it
is not too difficult to realize conditions that substantially
satisfy this statistical assumption.  The usual HBT analysis
requires, however, that the single-particle distribution be calculable from
an unsymmetrized two-particle source, and this condition is
much more difficult to fulfill.

   For resonance decays, we can ask whether the correlation
reflects the spatial propagation of the resonances before
they decay.
We find that the apparent source size extracted from
the quantum mechanically calculated correlation function
deviates strongly from the classical mean square size even for
resonance parameters and temperatures that suggest a classical
behavior. The off-shell propagation of resonances is important
and needs to be taken account of in the classical propagation
if extracted source sizes are to make sense.

\acknowledgements

This work has been supported in parts by the Department of Energy
under Grant DE-FG06-90ER40561
(G.B., M.H.),
by the National Science Foundation under
Grants PHY90-17077 and PHY89-04035 (P.D.), and
by the Alexander von Humboldt-Stiftung
(Feodor-Lynen Program) (M.H.).

\newpage

\begin{references}
\bibitem{Bo90}  D.\ Boal, C.K.\ Gelbke, and B.\ Jennings,
Rev.\ Mod.\ Phys.{\bf 62} (1990) 553.
\bibitem{Ak87} T.\ \AA{}kesson {\it et al.}, Z.\ Phys.\ {\bf C36} (1987) 517.
\bibitem{Se92}
D.\ Ferenc {\it et al.}, Nucl.\ Phys.\ {\bf A544} (1992) 531c,\\
P.\ Seyboth {\it et al.}, Nucl.\ Phys.\ {\bf A544} (1992) 293c.
\bibitem{Su93}  J.\ P.\ Sullivan, M. Berenguer, B.\ V.\ Jacak,
S.\ Pratt, M.\ Sarabura, J.\ Simon-Gillo, H.\ Sorge, and H. van Hecke,
 Phys.\ Rev.\ Lett.\ {\bf 70} (1993)
3000.
\bibitem{Hu93}  H.\ B\o{}ggild {\it et al.},
Phys.\ Lett.\ {\bf B302} (1993) 510.
\bibitem{Ak93} Y.\ Akiba {\it et al.}, Phys.\ Rev.\ Lett.\ {\bf 70} (1993)
1057.
\bibitem{Gy79} M.\ Gyulassy, S.K.\ Kaufmann, and L.W.\ Wilson,
 Phys.\ Rev.\ C {\bf 20} (1979) 2267.
\bibitem{Pr84} S.\ Pratt, Phys.\ Rev.\ Lett.\ {\bf 53} (1984) 1219.
\bibitem{Pr86} S.\ Pratt, Phys.\ Rev.\ D {\bf 33} (1986) 1314.
\bibitem{Be88}  G.\ Bertsch, M.\ Gong and M.\ Tohyama,
 Phys.\ Rev.\ C {\bf 37} (1988) 1896.
\bibitem{Gy89}  M.\ Gyulassy and S.\ Padula, Phys.\ Lett.\ {\bf B217}
(1989) 181.
\bibitem{Za84}  W.\ Zajc {\it et al.},
Phys.\ Rev.\ C {\bf 29} (1984) 2173.
\bibitem{KS92} K.\ Kadija and P. Seyboth, Phys.\ Lett. {\bf B287} (1992) 363.
\end{references}
\end{document}